\documentclass{ws-procs9x6}
\def\NIM{ {\it Nucl. Instrum. Methods} }
\def\NP {{\it Nucl. Phys.} }
\def\PL {{\it Phys. Lett.} }
\def\PRL{{\it Phys. Rev. Lett.}}
\def\PR {{\it Phys. Rev.} }

\def\be{\begin{equation}}
\def\ee{\end{equation}}
\def\bea{\begin{eqnarray}}
\def\eea{\end{eqnarray}}
\def\he3{$^3\vec{He}$}
\def\GEGM{${G_E/G_M} $}
\def\GEN{${G_E^n} $}
\def\GMN{${G_M^n} $}
\def\GENGMN{${G_E^n/G_M^n} $}
\def\GEP{${G_E^p} $}

\def\GEPGMP{${G_E^p/G_M^p} $}
\def\qsq{Q$^2$}
\def\gvsq{(GeV/c)$^2$}

\begin{document}

\title{Prospect for Measuring ${G_E^n}$ at High Momentum Transfers}

\author{B.~Wojtsekhowski}

\address{Physics Division, Thomas Jefferson National Accelerator Facility, \\
12000 Jefferson Avenue, Newport News, VA 23606, USA \\ E-mail: bogdanw@jlab.org}

\maketitle
\abstracts{Experiment E02-013, approved by PAC21, will measure 
the neutron electric form factor at \qsq~up to 3.4 \gvsq, which is twice
that achieved to date. The main features of the new experiment 
will be the use of the electron spectrometer BigBite, a large array of neutron 
detectors, and a polarized \he3~target.
We present the parameters and optimization of the experimental 
setup. A concept of an experiment for \GEN~where precision \GEP~data is 
used for calibration of the systematics of a Rosenbluth type measurement
is also discussed.}

\section{Introduction}

Elastic electron scattering, which in the one-photon approximation is characterized
by two form factors, is the simplest exclusive reaction on the nucleon. It provides 
important ingredients to our knowledge of nucleon structure. 
There are well-founded predictions of pQCD for the \qsq~dependence of the
form factors and their ratio in the limit of large momentum transfer \cite{br81}.
Predictions of a fundamental theory always attract substantial attention from 
experimentalists. Recent surprising results on \GEP~show that the ratio 
\GEPGMP~declines sharply as \qsq~increases, 
and therefore pQCD is not applicable up to 10 \gvsq. 
According to \cite{jo00,ga02} the electric and magnetic form factors behave differently, 
starting at \qsq~$\approx 1$ \gvsq. The same mechanisms causing this deviation should 
also be present in the neutron. It is an intriguing question, how the ratio 
\GENGMN~ develops in this \qsq~ regime, where confinement plays an important role.

\section{World data on \GEN}

\begin{figure}[t]
\begin{center}
\includegraphics[trim = 0mm 0mm 0mm 0mm, angle = 270, width=0.82 \textwidth]{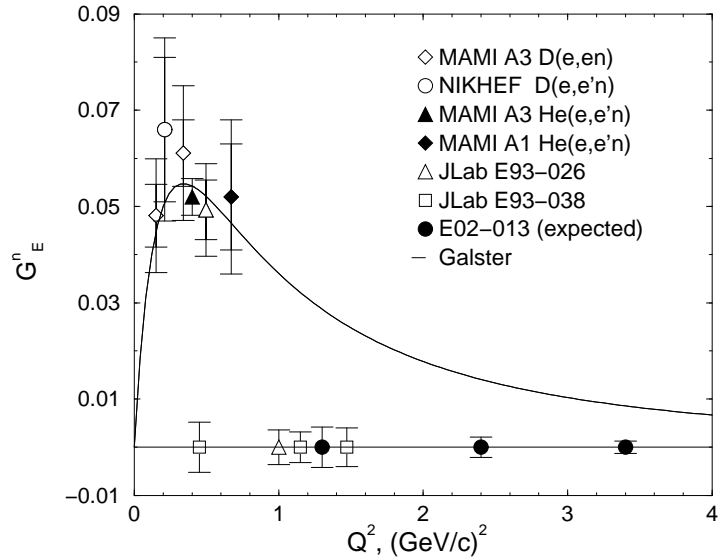}
\caption{ Recent data on \GEN~from double polarization experiments, the obtained 
accuracy and \qsq~of JLab experiments, and the expected accuracy of E02-013.}
\label{fig:w_data}
\end{center}
\end{figure}

The study of \GEN~ has been a priority in electromagnetic labs for the last 15 years.
Figure~\ref{fig:w_data} presents recent data \cite{pa99,os99,ro99,he99,zh01,go01} 
along with points representing the accuracy of JLab experiments \cite{da93,ma93} 
which have already collected data, and the expected statistical accuracy of
experiment E02-013. Presently published results can be fitted by the Galster 
approximation \cite{ga71}. The double polarization technique used in these 
experiments was introduced more than 20 years ago \cite{do69,ak74,ar81}. 
The experiments used a polarized electron beam and three different targets: 
unpolarized deuterium (together with a neutron polarimeter), 
polarized ND$_3$, and polarized $^3$He.

\section{Experiment E02-013}

The steady progress of the E93-028 \cite{da93} and the E93-026 \cite{ma93} 
experiments has made possible the accurate determination of \GEN~up to 1.47 \gvsq. 
The next step in \qsq~requires an experimental approach with much higher 
Figure-of-Merit (FOM).

In E02-013 \cite{ca02} we optimized the setup in several respects:
\begin{itemize}
\item the solid angle of the electron spectrometer, 
\item the neutron detector efficiency and the trigger logic,
\item the type of polarized target.
\end{itemize}

A recent addition in Hall A at JLab, the BigBite spectrometer developed by 
NIKHEF \cite{la98}, has a 76 msr solid angle for a 40 cm long target. 
We found that for the identification of quasi-elastic scattering, the momentum 
resolution of BigBite ($\approx$ 1\%) is sufficient for electron momenta up to 
1.5 GeV/c. The luminosity available with the \he3~target is about $10^{36}$~Hz/cm$^2$. 
According to our calculations it can be used with BigBite in spite of 
the direct view of the target by the detectors. 
Neutrons with kinetic energy above 1 GeV with which we have to deal at the proposed 
momentum
transfers, can be efficiently detected with a relatively high detector threshold, 
which allows to suppress background and is crucial for the operation at the expected 
luminosity, which is about a factor of 10 higher than used in a recent JLab 
experiment \cite{da93} with a polarized ND$_3$ target.

\begin{figure}[t]
\begin{center}
\includegraphics[trim = 0mm 0mm 0mm 0mm, angle = 90, width=0.82 \textwidth] {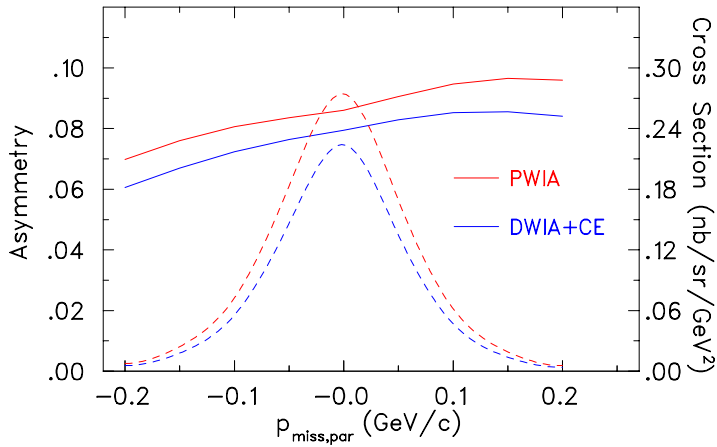} 
\caption{The GEA predictions for cross section and asymmetry in E02-013.
The error bars show the expected accuracy. 
For each \qsq~the asymmetry will be measured for three values of $p_{miss,perp}$.}
\label{fig:a_gen}
\end{center}
\end{figure}

In the last several years the theoretical development of the Generalized 
Eikonal Approximation (GEA) \cite{sa01} has provided a framework for taking into 
account nuclear effects in the extraction of \GEN~from the 
experimental asymmetry. The GEA prediction for the asymmetry as a function 
of the missing transverse momenta $p_{miss,perp}$ is shown in Fig.~\ref{fig:a_gen}. 
The GEA calculations and experimental data from JLab Hall B for the unpolarized reaction 
$^3$He(e,e'p) have demonstrated the dominance of quasi-elastic scattering 
at $p_{miss,perp}$ below 0.15 GeV/c, 
when a modest cut of 0.5 GeV/c is applied on $p_{miss,parallel}$.

Table~\ref{tab1} 
summarizes the contributions to the error budget 
for the highest \qsq~point.
For each \qsq~the measurement will be done with $\sim$14\% statistical accuracy 
for three intervals of $p_{miss,perp}$. As a result the systematics will be evaluated
by comparison of an experimental asymmetry and the GEA prediction vs $p_{miss,perp}$.
\begin{table}[th]
\tbl{The contributions to the error budget in 
\GEN~for the data point at \qsq=3.4~\gvsq. \vspace*{1pt}}
{\footnotesize
\begin{tabular}{|l|c|r|}
\hline
quantity    & expected value            & rel. uncertainty      \\ 
\hline
statistical error in raw asymmetry $A_{exp}$ & -0.0233 & 13.4\% \\
beam polarization $P_e$       & 0.75 & 3\%      \\
target polarization $P_{He}$  & 0.40 & 4\%      \\
neutron polarization $P_n$       & 0.86$\cdot P_{He}$& 2\%  \\
dilution factor $D$         & 0.94   &  3\%  \\
dilution factor $V$         & 0.91   &  4\%  \\
correction factor for $A_{parallel}$ & 0.94 & 1\%  \\
\GMN                                 & 0.057 & 5\%  \\
nuclear correction factor & 1.0 -- 0.85 & 5\%  \\ \hline
statistical error in \GEN~ & & 13.8\% \\
systematic error in \GEN~  & & 10.4\% \\ 
\hline
\end{tabular}\label{tab1} }
\end{table}
\section{Future considerations}

Experiment E02-013 is based on presently achieved parameters
of the \he3~target and the existing electron spectrometer. With 
additional developments the FOM of the experiment can be increased by 
a factor of 5 and a measurement of \GEN~will be feasible 
at \qsq~up to 5 \gvsq.

\subsection{Luminosity with the \he3~target} 

\begin{figure}[t]
\begin{center}
\includegraphics[trim = 0mm 0mm 0mm 0mm, width=0.82\textwidth] {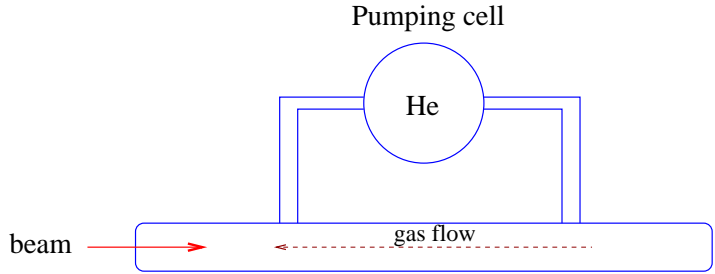} 
\caption{The target cell with two attachments to the pumping cell
which allow the gas flow.}
\label{fig:flow}
\end{center}
\end{figure}

The present configuration of the \he3~target has the highest FOM at 
a beam current of 12-15 $\mu$A, when the beam-induced depolarization time 
is on the order of 30 hours. The use of the higher beam current requires
a higher rate of polarizing and faster delivery of the polarized gas to 
the target cell.
Advances in solid-state laser technology have made available 100 and 
even 200 W light power suitable for polarizing Rb atoms.
Fig.~\ref{fig:flow} shows the target cell where the polarized gas flows 
through two tubes connecting the pumping and target cells. 
The flow will dramatically reduce the time for exchange of the polarized 
atoms between the pumping cell and the target cell. 

\subsection{High momentum spectrometer} 

\begin{figure}[t]
\begin{center}
\includegraphics[trim = 0mm 0mm -5mm 0mm,width=0.82 \textwidth]{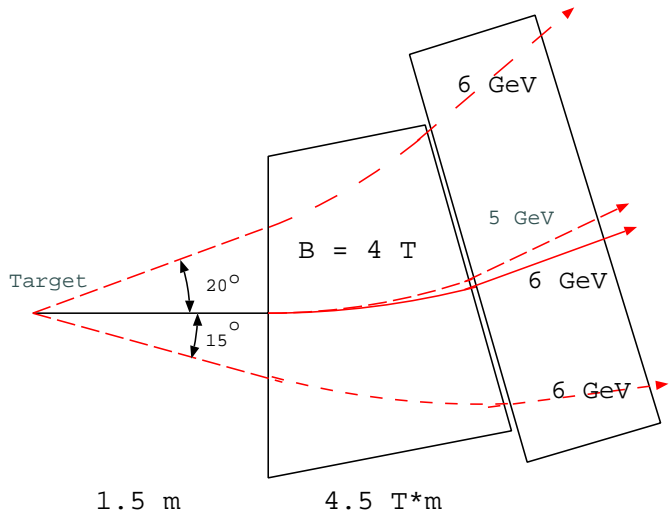}
\caption{A side view of the Super BigBite.}
\label{fig:sup_bb}
\end{center}
\end{figure}

The FOM of the experiment is approximately proportional to 
$E_f^2/E_i^2 \,=\, (E_i - Q^2/2M)^2/E_i^2$, where 
the $E_{i(f)}$ is the initial(final) electron energy. By using
a beam energy of 7.8 GeV it is possible to increase the FOM by a factor
of 2.7 in comparison to the plan in E02-013 \cite{ca02} for \qsq=3.4 \gvsq. 
It requires a new spectrometer for scattered electrons with a 
momentum 6 GeV/c and a solid angle of 75 msr. For \qsq= 5 \gvsq~the gain
of FOM is 3.4. The relative momentum
resolution should be of 0.5\% to keep a W resolution sufficient
for  identification of the quasi-elastic events. 
The base component of the spectrometer is a dipole magnet with a 
4.5 T$\cdot$m field integral and a 35 cm open gap. 
The scheme of a spectrometer based on such a dipole magnet is shown in 
Fig.~\ref{fig:sup_bb} \cite{ne02}. 
We call it Super BigBite. Its characteristics are similar to BigBite, 
but with the momentum range extended by a factor of 5-8.
As in the case of BigBite, the detector will be open to the target, 
so it can be used mainly with a polarized target luminosity.

\section{Rosenbluth approach}

\noindent
In the Rosenbluth method the form factors ratio $g \,=\,$\GEGM~ is obtained from two (or more) 
measurements at different beam energies at a fixed value of \qsq. 
The following equation is used to find $g$: 

\begin{equation}
g^2 \,=\, \tau \cdot \frac{ F^2_{\epsilon_1}\epsilon_2^{-1} - F^2_{\epsilon_2}\epsilon_1^{-1}}
{F^2_{\epsilon_2} - F^2_{\epsilon_1} } 
\label{eq:g_ratio}
\end{equation}
\noindent
where $F$ is the total form factor measured experimentally, 
$F^2 = \left( G_E^2 + \frac{\tau}{\epsilon}G_M^2 \right)/(1 + \tau) $, 
$\tau = \frac{Q^2}{4M^2_N}$ and $\epsilon$ is virtual photon polarization.
The uncertainty of $g$, which is growing with \qsq, 
can be estimated from the equation

\begin{equation}
\sigma(g^2) \,\approx\, \frac{\sigma(F^2_\epsilon)}{F^2_\epsilon} 
\frac{\sqrt{2} \cdot \tau }{\epsilon_1 \,-\, \epsilon_2}   
\label{eq:g_sigma}
\end{equation}
\noindent
where we neglect uncertainties in $\epsilon$ and $\tau$.  
The total form factor is calculated from the event 
rates and other parameters of the experiment as

\begin{equation}
 F^2 = N_{events}/[ I_{beam} \cdot d_{target} \cdot t_{DAQ} \cdot \sigma_{Mott} \cdot 
\Omega_e \cdot \eta_e]
\label{eq:exper}
\end{equation}

\noindent
Each of these experimental parameters - the beam current $I_{beam}$, the target density 
$d_{target}$, the data taking time $t_{DAQ}$, the Mott cross section $\sigma_{Mott}$, 
the detector solid angle $\Omega_e$, and the detection efficiency $\eta_e$ - is known 
with limited accuracy, which contributes to the systematics of the measurement. 
Some of them cancel in the calculation of the ratio $g$, 
because of the good stability of the target and detectors. A sufficiently accurate 
determination of the beam energy, the detector solid angle, and the scattering angle present 
a big challenge for the experiment. In the best case the overall systematic error is 
on the level of a few percent. By detecting the recoiling proton, as was suggested 
in LOI99-103 \cite{be99}, the acceptance of the detector can be excluded from the list of 
problems, because at a given value of the proton momentum the solid angle of the 
detector is fixed. Experiment E01-001\cite{ar01}, which used such an approach, 
recently took data in JLab Hall A. 

Quasi-elastic electron scattering from the deuteron $D(e,e'n)p$, with the 
ratio method suggested by Durand \cite{du59}, has been used 
for determination of the neutron magnetic form factor in recent experiments at 
Bonn \cite{br95}, Mainz \cite{ku02}, and JLab \cite{br94}.
The same reaction can be used for measurement of the ratio \GENGMN~even with less 
stringent requirements on the knowledge of the absolute neutron detection efficiency.
The small value of \GEN~made such measurement quite difficult; however, as we are proposing 
here, the problem can be solved  by using the complementary $D(e,e'p)n$ reaction for calibration
of the experiment. We will use the fact that in the \qsq~region of 5 \gvsq~the ratio of the 
proton form factors \GEPGMP~is already well known from JLab experiments \cite{jo00,ga02}. 
In a dedicated experiment the accuracy of $g_p$ can be improved to the level of 2-3\%.

The proposed scheme will use the magnetic spectrometer as an electron arm and a
non-magnetic detector as a hadron arm. The last one will consist of a large array of plastic
scintillators and veto detectors. At a few \gvsq~momentum transfer the kinetic energy of 
the recoiling nucleon is above 1 GeV and proton and neutron interactions with the 
detector are similar (nuclear interaction dominates).
The neutron detection efficiency of different measurements will be similar 
to each other because of equal kinetic energy of the neutron in both measurements.
Most of the remaining variations of the detector efficiency and solid angle 
will affect in the same way the complementary reaction $D(e,e'p)n$. 

The ratio $F^2_{\epsilon_2}/F^2_{\epsilon_1}$, which defines as the value of $g_n$,
can be expressed in the proposed experiment as

\begin{equation}
 \left( \frac{F^n_{\epsilon_2}}{F^n_{\epsilon_1}} \right)^2 \,=\,
 \left( \frac{F^p_{\epsilon_2}}{F^p_{\epsilon_1}} \right)^2 \cdot
 \frac{N^{e,e'n}_2}{N^{e,e'n}_1} \cdot  \frac{N^{e,e'p}_1}{N^{e,e'p}_2} \cdot
 \frac{\Omega^n_{\epsilon_2}}{\Omega^n_{\epsilon_1}} \cdot 
 \frac{\Omega^p_{\epsilon_1}}{\Omega^p_{\epsilon_2}} \cdot
 \frac{\eta^n_{\epsilon_2}}{\eta^n_{\epsilon_1}} \cdot 
 \frac{\eta^p_{\epsilon_1}}{\eta^p_{\epsilon_2}}
\label{eq:n_p_g}
\end{equation}

Several parameters such as the beam current, the electron-arm solid angle and efficiency,
the Mott cross section, the data taking time, and the target parameters all cancel out 
from the final ratio of the form factors at two different values of $\epsilon$.
The remaining parameters are the neutron/proton detector solid angle $\Omega$ 
and efficiency $\eta$, whose variations for different $\epsilon$ need to be controlled.

For the proposed non-magnetic detector at large nucleon energy the neutron 
and proton detector efficiency will be almost equally affected by any change
of rates and drifts of detector parameters, so it will be compensated. The detector solid
angle is defined by the detector size. It can be well controlled and small changes
will be the same for both proton and neutron channels.

The prospect of the Rosenbluth approach for a measurement of \GEN~depends on the high
rate capability of the neutron detector. The potential FOM is higher than that 
possible in the double polarization approach by a factor 10-20, when it operates 
at a luminosity of $10^{38}$~Hz/cm$^2$.
Experiment E93-038\cite{ma93}, which was done at a similar luminosity, developed
the appropriate techniques for background reduction.

\section*{Conclusion}

The experimental field of neutron electromagnetic form factors made very good progress
in recent years. The present frontier for \GEN~is $Q^2$ above 2 \gvsq.  
JLab experiment E02-013 will do the measurement of \GEN~up to $Q^2 = 3.4$ \gvsq.
There are possibilities of the further enhancements of the luminosity and polarization 
of the polarized \he3 target. The Rosenbluth approach may also be revived by using 
calibration on the proton \GEPGMP~ratio. 

\section*{Acknowledgments}

It is pleasure to thank G.~Cates, B.~Reitz, and K.~McCormick for
collaboration in developing the E02-013 experiment.
I would like also to acknowledge the crucial contributions made 
by P. Degterenko, V. Nelyubin, M. Sargsyan, and by members
of the Hall A collaboration. Discussions with W.~Brooks, K.~de Jager,
and B.~Mecking on the possibility of the Rosenbluth approach
for the \GEN~measurement are greatly appreciated. The Southeastern Universities 
Research Association operates the Thomas Jefferson National Accelerator
Facility under Department of Energy contract DE-AC05-84ER40150.

\end{document}